\documentclass[12pt]{article}
\usepackage{latexsym,epsfig,graphicx,amsmath,amssymb,amscd,multirow}
\usepackage{chicago,psfrag,paralist,dsfont,url,comment}
\usepackage{subfig}
\usepackage{graphicx}
\usepackage[titletoc]{appendix}

\usepackage[american]{babel}

\newcommand{\bx}{\boldsymbol{x}}

\newcommand{\bw}{\boldsymbol{w}}
\newcommand{\bv}{\boldsymbol{v}}

\newcommand{\zerovec}{\boldsymbol{0}}

\newcommand{\RMSE}{\textrm{RMSE}}

\newcommand{\GAMMA}{\textrm{Gamma}}
\newcommand{\BETA}{\textrm{Beta}}

\newcommand{\tP}{\text{P}}
\newcommand{\tL}{\text{L}}

\newcommand{\NOR}{\mathcal{N}}
\newcommand{\FDIST}{\mathcal{F}}

\newcommand{\btheta}{\boldsymbol{\theta}}
\newcommand{\bt}{\boldsymbol{t}}
\newcommand{\by}{\boldsymbol{y}}

\newcommand{\bnu}{\boldsymbol{\nu}}

\newcommand{\nurL}{\nu_{r}^{\text{L}}}
\newcommand{\nucL}{\nu_{c}^{\text{L}}}

\newcommand{\tran}{^\top}

\newtheorem{preprop}{Proposition}
{\begin{preprop}\upshape}
{\end{preprop}}

\newtheorem{pretheo}{Theorem}
\newenvironment{theorem}%
{\begin{pretheo}\upshape}
{\end{pretheo}}

\begin{document}


\title{\vspace{-4ex} Modeling Spatially Correlated Failure-time Data Under Two Distance Functions with an Application to Titan GPU Data}


\author{
Jared M. Clark$^1$, Jie Min$^2$, Yueyao Wang$^3$\footnote{Corresponding Author. Email: yueyaowang@mail.zjgsu.edu.cn}, \\ Yili Hong$^1$, and George Ostrouchov$^4$ \\
{\small $^1$Department of Statistics, Virginia Tech, Blacksburg, VA 24061, USA}\\
{\small $^2$Department of Mathematics and Statistics, University of South Florida,}\\[-0.7ex]
{\small Tampa, FL 33620, USA}\\
{\small $^3$School of Statistics and Mathematics, Zhejiang Gongshang University,}\\[-0.7ex]
{\small Hangzhou, Zhejiang 310018, China}\\
{\small $^4$Department of Business Analytics and Statistics, University of Tennessee,}\\[-0.7ex]
{\small Knoxville, TN 37996, USA}
}

\date{}
\maketitle
\vspace{-5ex}	
\begin{abstract}
One common approach to statistical analysis of spatially correlated data relies on defining a correlation structure based solely on unknown parameters and the physical distance between the locations of observed values. However, some data have a complex spatial structure that cannot be adequately described with the physical distance alone. In this work, the spatial failure-time data of focus contains information on GPUs that are connected through a network fabric topology that differs from their physical layout and that is expected to introduce additional correlations. The proposed lifetime regression model includes random effects capturing the dependency due to physical location as well as random effects explaining the dependency due to logical connections between GPUs. The analysis of this GPU dataset serves as an example of models with multiple spatial random effects and the ideas presented can be extended to other applications with complex spatial structures. A Bayesian modeling scheme is recommended for this class of analyses. The examples in this work use the software package, Stan, to produce Markov chain Monte Carlo draws for parameter estimation. This modeling effort is validated through simulation which demonstrates accuracy in statistical inference. We also apply the developed framework to the large-scale Titan GPU failure time data.

\textbf{Key Words:}  Accelerated failure time model; GPU lifetime; Logical connections; Physical connections; Spatial survival data; Supercomputer reliability.

\end{abstract}

\newpage
	
\section{Introduction}\label{sec:introduction}
\subsection{Motivation} 
Spatially correlated failure-time data has received a great deal of attention in many areas including clinical trials, bio-statistics, and engineering. In the literature, spatial effects are typically incorporated in the model as random effects with spatial correlations. Usually, the failure-time data in adjacent regions have higher correlation, because closer locations share similar environmental characteristics. Typically spatial correlations are assumed to be a function of the physical distance between two locations.

In modern engineering fields, however, products often come with more complex spatial structures, where physical distance alone cannot explain the entirety of the spatial correlation structure. High-performance computing (HPC) is one example with such a complex design structure. HPC sees a massive amount of regularly arranged graphical processing units (GPUs) linked through some logical connections so that the system has high level performance for large-scale computing. The GPU nodes are installed in cages and placed in multi-layer cabinets. Many cabinets form a rectangular layout in a server room and there is a network fabric that connects GPU nodes to form a logical structure.

The architecture of HPC clusters contains both physical and logical structures so when modeling the GPU failures, there are two types of spatial distance to consider. One is the physical distance between cluster cabinets and the other is defined by the number of direct logical connections to traverse for communication. The physical distance is thought to affect GPU failure through heat dissipation in the server room, as overheating is one major reason of GPU failure. It is expected that GPU nodes closer to each other physically are under similar heat dynamics so they will experience similar failure patterns. While for logical connections, GPU failure is affected through job scheduling since jobs are allocated in mostly contiguous logical blocks for faster communication over the network fabric. It is expected that the lifetime of two GPU's with fewer connections between them will have higher correlation. It is unknown how these two types of distance impact the GPU lifetime. With the knowledge the temperature is related to rate of failure, modeling these correlation sources has some motivation. Energy input into the system is controlled by job scheduling while energy dissipation is primarily a spatial process. Therefore, it is desirable to consider both dependencies due to physical location as well as those due to the number of logical connections between GPUs in the model.

In this study, we are motivated by the Cray XK7 Titan supercomputer data~\shortcite{ostrouchov2020gpu} to investigate spatial correlation using different types of distance functions. The Titan supercomputer system was the worldwide leading computer system for some time and played an important role in critical computing missions. For Titan supercomputers, more than 30{,}000 GPUs are placed in an 8 by 25 grid of cabinets in the server room. The failure event time, unit status and physical location of Titan GPUs were recorded over a 7 year service period, providing us an opportunity to study the spatial structure and its effect on the GPU reliability. The details of the dataset are introduced in Section~\ref{sec:data_model}. Based on the HPC spatial structure, we propose to extend the traditional lifetime regression model by including two sets of spatial random effects: one to capture the correlation inherited in the physical structure of Titan supercomputers and the other to capture the correlation induced by the logic connections.

\subsection{Related Literature and Contribution of This Work}
During recent years, modeling spatially correlated survival data has received considerable attention in the literature. For example, \citeN{henderson2002modeling} augmented the Cox model with spatial frailty terms using a hierarchical approach and incorporated the spatial correlation in the mean frailty effects in analyzing leukemia survival data. \citeN{hennerfeind2006geoadditive} added spatial effects to the hazard function of the Cox model and explored multiple types of priors for the spatial effects. \citeN{li2002modeling} and  \shortciteN{li2015survival} added normally distributed random effects with multiple correlation structures to the hazard functions of Cox models. Based on \citeN{li2002modeling}, \citeN{motarjem2020geostatistical} further proposed a geostatistical spatial survival approach for Cox models. Using a Bayesian approach,  \citeN{zhang2011bayesian} added normally distributed spatial random effects with a conditional autoregressive (CAR) structure to the accelerated failure-time (AFT) model in analyzing prostate cancer survival time. \citeN{wang2016bayesian} combined the random effects with the CAR prior and mixture AFT models. \citeN{hanson2012bayesian} also considered random effects with the CAR prior and proposed a stratified proportional hazards model with spatial frailty terms. \shortciteN{pan2014bayesian} explored the combination of spatial random effects with the CAR prior and Cox model for interval censored data and proposed a specific MCMC algorithm for the model using latent variables.  \citeN{zhou2017generalized} proposed an AFT spatial frailty model with a CAR prior that can be applied to arbitrary censored data.  \citeN{zhou2018unified} further introduced a statistical framework to analyze arbitrarily censored spatial survival data with Gaussian random effects and multiple popular semi-parametric models. \shortciteN{motarjem2019bayesian} proposed adding spatial random effects with a non-Gaussian CAR structure to the Cox model when the survival data is skewed or has a heavy tail distribution. \citeN{geng2021bayesian} introduced a proportional hazards model that captures the spatial pattern using both regression coefficients and baseline hazard functions.

Furthermore, modeling spatial correlation with the existence of competing risks is also considered in the literature.  \shortciteN{hesam2018spatial}, \citeN{momenyan2021joint}, and \citeN{momenyan2022competing} added spatial random effects into the hazard functions of cause-specific and subdistribution competing risks models in analyzing cancer and HIV/AIDS data. \shortciteN{min2023spatially} combined spatial random effects and the AFT model with the presence of competing risks in analyzing the Titan GPU data. Additionally, spatial temporal survival models are developed to analyze the infection risk of wildlife disease (\shortciteNP{yao2023bayesian}), the survival time of prostate cancer (\shortciteNP{wang2024spatial}), and the recurrent failures of telecommunication base station systems (\shortciteNP{Wu02012024}).

To the best of our knowledge, all the existing methods only consider one type of spatial correlation. However, using a single type of distance function and spatial correlation function can be inadequate to capture the patterns in data. For example, in analyzing the lifetime of GPUs inside the Titan supercomputer,  \shortciteN{ostrouchov2020gpu} and \shortciteN{min2023spatially} suggested that the spatial correlation comes from both physical GPU locations and logical GPU locations, indicating that different spatial structures should be considered based on the two sources. In addition, the physical and logical locations are not unique for Titan, but commonly seen in other supercomputer architectures. To solve this problem, we propose a model that is able to capture spatial correlations from different sources with different distance functions and spatial covariance structures. In particular, we incorporate both Euclidean distance and circle distance in the proposed model. In addressing the complex spatial structure in this work, it becomes necessary to define distance on a torus. \citeN{gneiting2013strictly} and \citeN{porcu2016spatio} suggested that spatial correlation functions developed based on the Euclidean distance need to be modified to maintain their positive definiteness when distances on the sphere are used. The two papers also developed spatial correlation functions that can be used for circle distances. In this paper, we further propose a separable spatial correlation function for distance on the torus, which has yet to be done in the literature. Our statistical analysis of Titan GPU lifetime data provides insights into GPU failure mechanisms and sheds light on the future design of supercomputers.

\subsection{Overview}

The rest of the paper has the following organization. Section~\ref{sec:data_model} introduces the relevant notation and presents a novel mixed-effects model for time-to-event data that incorporates multiple sets of random effects. This section also describes the Cray XK7 Titan GPU dataset, which serves as a motivating example. Section~\ref{sec:Bayesian.inference} provides a brief review of the Bayesian framework and discusses its relevance to the data analysis. Section~\ref{sec:simulation.study} presents a simulation study to evaluate the performance of the proposed model. Section~\ref{sec:data_analysis} details the analysis of the Titan GPU dataset and discusses the results. Finally, Section~\ref{sec:conclusion} offers concluding remarks and identifies directions for future research.

\section{Data and Model}\label{sec:data_model}
In this section, we first introduce the data notation as well as notation for a general mixed effects model with two sets of spatial random effects in Section~\ref{sec:general data notation}. Then in Section~\ref{sec:model_with_gpu}, we describe in detail the physical and logical spatial components associated with the GPU data. Furthermore, in Section~\ref{sec:dist.functions}, we justify the positive definiteness of the correlation matrix with the proposed distance functions.
\subsection{Data Notation and the General Model}\label{sec:general data notation}

To handle reliability data with complex spatial structures, a mixed effects model that extends the familiar accelerated failure-time (AFT) model is proposed. This model is general enough to accommodate multiple parametric families of distributions as well as multiple correlation structures. Suppose there are $n$ units in the dataset and $p$ covariates of interest. Let $\boldsymbol{t} = (t_1, t_2, \dots, t_n)\tran$ be the vector of event times with $t_i$ the observed failure/censoring time of unit $i$,
 and $\delta_i$ the corresponding failure indicator such that $\delta_i = 1$ indicates a failed unit. That is we record $t_i$ as the underlying failure-time $\tilde{t}_i$ if observation $i$ is an event, and we record $t_i$ as the censoring time otherwise. Additionally, let $\boldsymbol{x}_{i} = ( 1,  x_{1i},  x_{2i},  \dots,  x_{pi} )\tran$ be the vector of explanatory information for the $i$th unit. Without loss of generality, assume the event time of each unit is correlated with the others where the correlation arises from two sources. Then the proposed model takes the following form:
\begin{align}\label{eq:generalModel}
\by = X\boldsymbol{\beta} + Z_{\bv}\boldsymbol{v}+Z_{\bw}\boldsymbol{w} + \boldsymbol{\epsilon},
\end{align}
where $\by=\left(\log(\tilde{t}_1), \log(\tilde{t}_2), \dots, \log(\tilde{t}_n)\right)\tran$ is a length $n$ vector with log failure-times. Note that there are typically some failure-times unobserved due to censoring. Here, $X = \left(\bx_1, \bx_2, \dots, \bx_n\right)\tran$ is the $n \times (p+1)$ model matrix containing explanatory information, and $\boldsymbol{\beta} =(\beta_0, \beta_1, \beta_2, \dots, \beta_p )\tran$ is a vector of linear coefficient parameters associated with the fixed effects. The matrices $Z_{\bv}$ and $Z_{\bw}$ identify the location of observations under two spatial structures. We use $\boldsymbol{v}, \boldsymbol{w}$ to denote the random vectors that contain the two types of spatial effects, and $\boldsymbol{\epsilon}$ to denote the noise vector that captures the information not explained by the fixed and spatial random effects. The detailed expressions of location matrices  $Z_{\bv}, Z_{\bw}$ and distributional assumptions of $\boldsymbol{v}, \boldsymbol{w}$ will be introduced shortly.

The noise terms $\epsilon_i$, for $i = 1, 2, \dots, n$, are typically assumed to be independent and identically distributed, following a distribution from a location-scale family, which means $\epsilon_i \sim \FDIST(0, \sigma)$ and $\boldsymbol{\epsilon} = (\epsilon_1, \epsilon_2, \dots, \epsilon_n)\tran$. Denote the variance-covariance matrix for $\boldsymbol{\epsilon}$ as $\Sigma_\epsilon$. The relationship between $\Sigma_\epsilon$ and the scale parameter may not be trivial if we do not have normality. The transformation of event time to the log scale in model~(\ref{eq:generalModel}) removes restrictions on the support of $\epsilon_i, i = 1,\dots, n$. Usually $\FDIST$ is assumed to be a normal (NOR) or smallest extreme value (SEV) distribution. These assumptions translate to a log-normal or Weibull assumption on the observed scale of the event-time data.

This model differs from typical accelerated failure-time models with the inclusion of multiple sets of random effects, $\boldsymbol{v}$ and $\boldsymbol{w}$. The random effects are modeled in the following manner:
\[ \boldsymbol{v} \sim \NOR(\boldsymbol{0}, \Sigma_{\bv}), \: \boldsymbol{w} \sim \NOR(\boldsymbol{0}, \Sigma_{\bw}), \]
where
\begin{align*}
 \boldsymbol{v} = (v_1, v_2, \dots, v_m )\tran, \: \boldsymbol{w} = (w_1, w_2, \dots, w_m)\tran.
\end{align*}
Let $j=1,2,\dots,m$  be the index for the location and $m$ be the total number of locations. In this case, $v_j$ is the effect at location $j$ due to the first spatial component, while  $w_j$ is the effect at location $j$ due to the second spatial component. Note that in some cases it might be convenient to use different labeling systems for the two spatial components, in which case we would introduce a new index, $l \in \{1,2,\dots,m\}$.

One key assumption of this model is that $\boldsymbol{v}$ and $\boldsymbol{w}$ are independent random vectors. This is to say that any subset of elements in $\boldsymbol{v}$, denoted by $\boldsymbol{v}^{\ast}$, is independent of any subset of elements in $\boldsymbol{w}$, denoted by $\boldsymbol{w}^{\ast}$. Equivalently,
\begin{align*}
 f_{\bv^{\ast},\bw^{\ast}}(\boldsymbol{v}^{\ast},\boldsymbol{w}^{\ast}) = f_{\bv^{\ast}}(\boldsymbol{v}^{\ast})f_{\bw^{\ast}}(\boldsymbol{w}^{\ast}).
\end{align*}
This assumption indicates that while an element of $\boldsymbol{v}$ may be informative of other elements in $\boldsymbol{v}$, it carries no information about $\boldsymbol{w}$. Likewise, the independence assumption makes modeling the complex spatial structure feasible and is supported in our application by independence of energy input structure of job scheduling from energy dissipation structure of ambient air temperature.

We also have the two $n \times m$ matrices, $Z_{\bv}$ and $Z_{\bw}$. If we haven't introduced any relabeling of the locations for the two spatial components then $Z_{\bv}=Z_{\bw}$. Now, $\boldsymbol{z}_{vi}$ is an $m$-vector containing $(m-1)$ zeros. The remaining element is a one located in the $j$th spot indicating that unit $i$ is located at location $j$, in terms of the first spatial effect. The vector $\boldsymbol{z}_{wi}$ is defined similarly, as a vector of zeros and a single one, but indicates the position in terms of the second spatial effect. Then,
\[ Z_{\bv} = \begin{pmatrix} \boldsymbol{z}_{v1}\tran \\ \boldsymbol{z}_{v2}\tran \\ \vdots \\ \boldsymbol{z}_{vn} \tran\end{pmatrix}, Z_{\bw} = \begin{pmatrix} \boldsymbol{z}_{w1}\tran \\ \boldsymbol{z}_{w2}\tran \\ \vdots \\ \boldsymbol{z}_{wn} \tran\end{pmatrix}. \]
Without loss of generality, we can assume that the observations are ordered by location and focus on the first spatial structure. Let $n_j$ denote the number of observations in location $j$, then the $Z_{\bv}$ matrix maps each observation to its spatial location and it can be expressed as,
\[ Z_{\bv} = \begin{pmatrix} \boldsymbol{1}_{n_1} & \boldsymbol{0}_{n_1} & \hdots & \boldsymbol{0}_{n_1} \\
\boldsymbol{0}_{n_2} & \boldsymbol{1}_{n_2} & \hdots & \boldsymbol{0}_{n_2} \\
\vdots & \vdots & \ddots & \vdots \\
\boldsymbol{0}_{n_m} & \boldsymbol{0}_{n_m} & \hdots & \boldsymbol{1}_{n_m} \end{pmatrix} .  \]

Now we are able to address the covariance structure of the model response $\by$ in \eqref{eq:generalModel}. The covariance matrices can take on many forms depending on the structure of the data. Note that both $\Sigma_{\bv}$ and $\Sigma_{\bw}$ should be positive definite. As a result of our independence assumptions, the covariance matrix of $\by$ is denoted as $\Sigma$ and takes the following form:
\begin{align*}
\Sigma = Z_{\bv}\Sigma_{\bv}Z_{\bv}\tran + Z_{\bw}\Sigma_{\bw}Z_{\bw}\tran + \Sigma_\epsilon\,.
\end{align*}
As the sum of positive definite matrices, $\Sigma$ is a positive definite matrix.

In order to reduce the number of parameters to be estimated, it is suggested that a correlation function is chosen to define the dependence structure. The powered exponential structure is widely used due to it's large degree of flexibility. Consider a two dimensional space where the two axes are denoted as $r$ and $c$ respectively. Then for the random effects $v_j$ and $v_{j'}$, that correspond to locations $(r_j,c_j), (r_{j'},c_{j'})$, the power exponential correlation function under the first spatial structure is:
\begin{align}
\rho[(r_j,c_j), (r_{j'},c_{j'})] = \exp\Big\{ - \Big[\frac{d_{r}(r_j,r_{j'})}{\nu_{r}}\Big]^\kappa - \Big [\frac{d_{c}(c_j,c_{j'})}{\nu_{c}}\Big]^\kappa\Big\} .
 \end{align}
In this case, $d_{r}(\cdot,\cdot)$ and  $d_{c}(\cdot,\cdot)$ are appropriate one-dimensional distance metrics. The $\nu_{r}$ and $\nu_{c}$ are length-scale parameters in each direction that control the correlation decay rate and are bounded below by 0. The $\kappa$ parameter is shared across all dimensions and adds flexibility to the distance space. This parameter is confined to an interval which ensures positive-definiteness, though the bounding interval is dependent on the chosen distance function. Note that if circle distance is used, $\kappa$ is bounded above by 1, while $\kappa$ is bounded above by 2 for Euclidean distance. If $\sigma_{\bv}$ is the marginal variance of the random effects $\boldsymbol{v}$ under the first spatial component and $R_{\bv} = (\rho_{st}), s = 1, \dots, m, t = 1, \dots, m$, computes the $m \times m$ correlation matrix for $\boldsymbol{v}$, then:
\begin{align*}
 \Sigma_{\bv} = \sigma_{\bv}^2 R_{\bv},
\end{align*}
is the variance-covariance matrix for $\boldsymbol{v}$. Now $\Sigma_{\bw}$, for the random effect $\boldsymbol{w}$, can be defined in a similar manner. Note that the distance functions can differ from those used in defining $R_{\bv}$. It would also be possible to define $R_{\bv}$ with a different correlation function, however we assume that the powered exponential structure is flexible enough for this work.

\subsection{Physical and Logical Spatial Correlation with the GPU Data}\label{sec:model_with_gpu}

As introduced in Section~\ref{sec:introduction}, the complex physical and logical connections between GPU nodes motivate the mixed effects model with two sets of spatial random effects. In this section, we introduce the physical and logical spatial components and present how the distance and correlation functions are defined with the GPU data.

\begin{figure}
	\centering
	\begin{tabular}{cc}
		\includegraphics[width=0.35\linewidth]{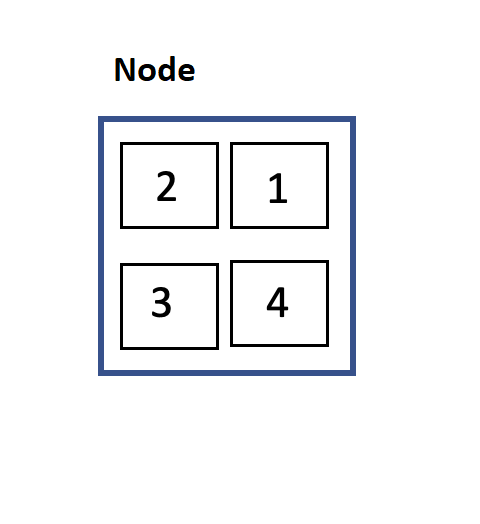}
		&\includegraphics[width=0.5\linewidth]{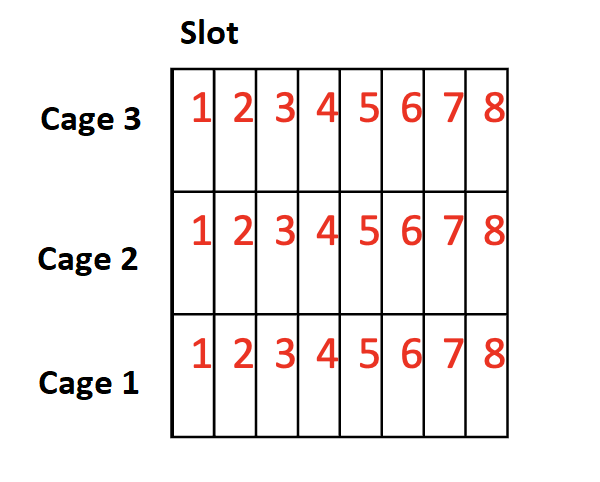}\\
		(a) 4 GPU nodes in each slot& (b) 8 slots in each cage and 3 cages in each cabinet \\
	\end{tabular}
	\caption{The physical organization of Titan supercomputer inside a cabinet.}\label{fig:gpu}
\end{figure}

GPU clusters inside supercomputers such as Cray XK7 Titan are organized within cabinets. For the Titan supercomputer, GPUs are arranged by cage (3 levels), slot (8 levels), and node (4 levels) within each cabinet. In the context of our mixed effects model, these are treated as fixed effects. It is believed that GPUs located higher within a cabinet will experience failures at a higher rate due the higher temperatures toward the top of the cabinets. These GPUs are located farthest from the cooling unit at the bottom of the cabinet. Cage 3 occupies the upper third of the cabinet while cage 1 occupies the lower third, with cage 2 in the center. Slots are placed side-by-side within a cage, indicating the horizontal placement of a GPU within the cabinet. Within a cage and slot, nodes 1 and 2 are located above nodes 3 and 4.  Figure~\ref{fig:gpu} shows the physical organization of GPU nodes, slots and cages within a cabinet. Cage, slot and node are used to define the columns of the model matrix, $X$. Cage 3, slot 8 and node 4 will serve as baseline categories, meaning they do not have associated columns within $X$. The first column in $X$ is a column of ones which incorporates the intercept, $\beta_0$ through matrix multiplication. The remaining $p = 12$ columns in $X$ are indicator variables which can be identified by their corresponding coefficient parameters.

The cabinets for the Titan GPU cluster were spread across 8 rows and 25 columns for a total of $m=200$ unique locations. The spatial structure of the data is based solely on these 200 locations and not the location of a GPU within a cabinet. Let $j = 1, 2, \dots, m$ be the index of locations, $r_j$ and $c_j$ denote the row and column for location $j$. Additionally, let $n_r$ be the total number of rows and $n_c$ be the total number of columns.

The first component of the spatial structure depends on the physical layout of cabinets within the server room.  The random effects associated with the physical location of cabinets will be contained in the $m \times 1$ vector $\boldsymbol{v}$. Once again, temperature is the motivating reason for believing that these effects should be included in the model. However, now we are considering a more ``global'' room temperature. GPUs located close to each other within the room will experience similar environmental factors which will induce correlation. The two dimensions of physical distance are defined by the rows and columns. We use absolute distance in each direction to describe the distance between two locations $(r_s, c_s)$ and $(r_t, c_t)$ inside the physical spatial correlation, where the superscript $(\cdot)^{\text{P}}$ indicates the notation is associated with the physical spatial component:
\begin{align}
 d^{\text{P}}_{r}(r_s,r_t) = |r_s-r_t|, \; d^{\text{P}}_{c}(c_s,c_t) = |c_s-c_t|.
\end{align}
Now, the correlation between the physical random effects at locations $(r_s, c_s)$ and $(r_t, c_t)$ is defined by the power exponential function $\rho^{\text{P}}_{st} = \exp\left[-b^\tP_{st}(\nu_{\bv}, \kappa_{\bv})\right]$ as in Section~\ref{sec:general data notation}, where
\begin{align}\label{eqn:physical.dist}
	b^\tP_{st}(\nu_{\bv}, \kappa_{\bv})&= \left(\frac{d_{r}^\tP}{\nu_{r}^\tP}\right)^{\kappa_{\bv}}+
	\left(\frac{d_{c}^\tP}{\nu_{c}^\tP}\right)^{\kappa_{\bv}}  =
	\left(\frac{|r_s-r_l|}{\nu_{r}^\tP}\right)^{\kappa_{\bv}}+
	\left(\frac{|c_s-c_l|}{\nu_{c}^\tP}\right)^{\kappa_{\bv}},
\end{align}
$\bnu_{\bv}=(\nu_{r}^\tP, \nu_{c}^\tP)\tran$ and $0<\kappa_{\bv}\leq 2$.

What are referred to as ``logical'' connections define the second component of the spatial structure. Cables connect the cabinets together which allows for communication between the GPUs in different cabinets. For the Titan supercomputer, every other cabinet is directly connected along both column and row directions in a so-called folded torus topology to minimize the maximum cable length \cite{ezell2013understanding}. It is believed that there will be some dependency in the failure-times of GPUs based on how many logical connections are between their respective cabinets.

\begin{figure}
	\centering
	\includegraphics[width=.9\textwidth]{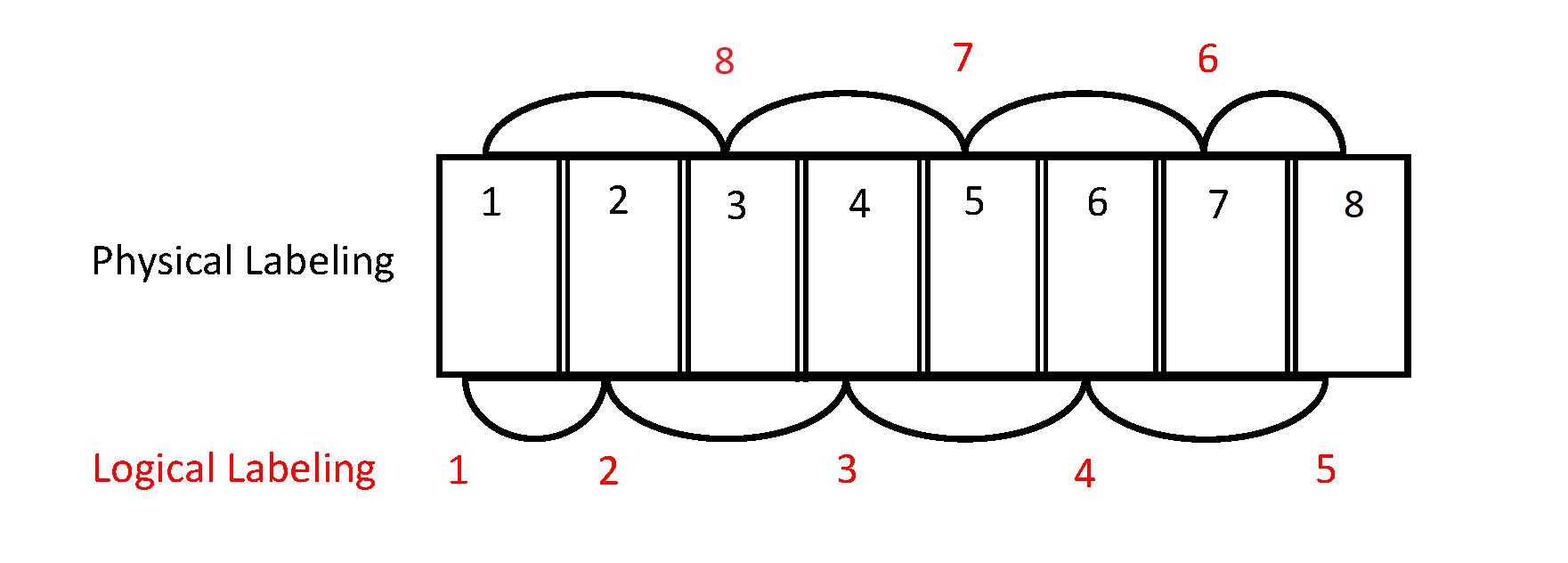}
	\caption{Illustration of the logical connections among cabinets in the same column.}\label{fig:logcial connections}
\end{figure}

\begin{figure}
	\centering
	\includegraphics[width=.65\textwidth]{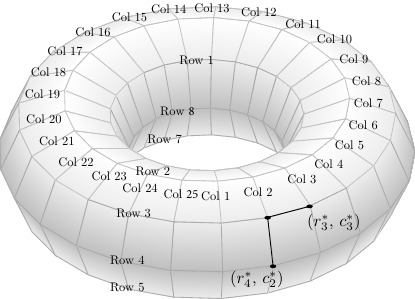}
	\caption{Illustration of distance on torus.}\label{fig:torus.dist.plot}
\end{figure}

However, a Euclidean-like approach is inappropriate for measuring the logical distance between cabinets. The primary difficulty arises in that the cable connections between cabinets form a circuit (both in terms of rows and columns). See Figure \ref{fig:logcial connections} for a depiction of the cable circuit in the rows. We introduce a relabeling scheme for the logical distance, $(r_j^{\ast},c_j^{\ast})$, which reflects the position of location $j$ within the circuit. We then use the following distance functions with the superscript $(\cdot)^\tL$ indicating the logical spatial component:
\begin{align}
 d^{\text{L}}_{r}(r_s^{\ast},r_t^{\ast}) &= \min\{|r_s^{\ast}-r_t^{\ast}|, n_r- |r_s^{\ast}-r_t^{\ast}|\} ,\;\\ \nonumber
d^{\text{L}}_{c}(c_s^{\ast},c_t^{\ast}) &= \min\{|c_s^{\ast}-c_t^{\ast}|, n_c- |c_s^{\ast}-c_t^{\ast}| \} .
\end{align}
In both the row and column dimensions we are working with locations on a circle. Note that the Cartesian product of two circles is a torus (a doughnut-shaped surface). Therefore, it has become necessary to define distance on a torus. Figure~\ref{fig:torus.dist.plot} illustrates the defined distance metric on a torus.

\begin{figure}
	\centering
	\includegraphics[width=0.95\textwidth]{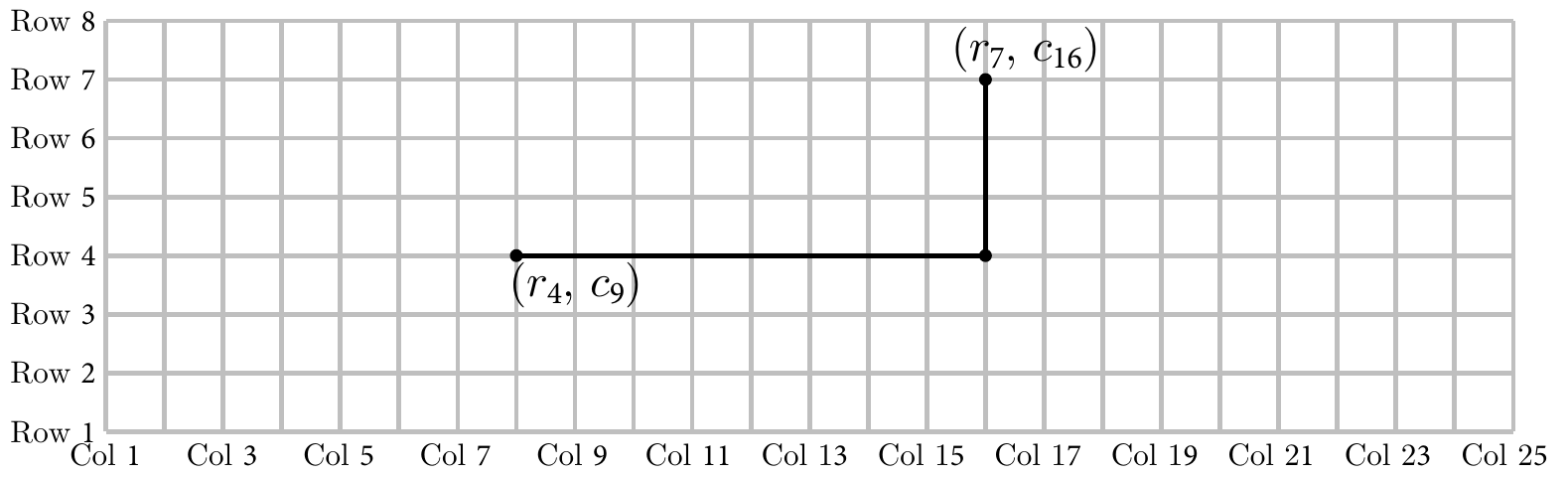}
	\caption{Illustration of the physical distance.}\label{fig:phy.dist.plot}
\end{figure}

Similarly, the correlation between the logical random effects at locations $(r_s^{\ast}, c^{\ast}_s)$ and $(r^{\ast}_t, c^{\ast}_t)$ is defined by $\rho^{\text{L}}_{st} = \exp\left[-b^\tL_{sl}(\nu_{\bw}, \kappa_{\bw})\right]$:
\begin{align}\label{eqn:logical.dist}
	b^\tL_{sl}(\nu_{\bw}, \kappa_{\bw})&=
	\left(\frac{d_{r}^\tL}{\nu_{r}^\tL}\right)^{\kappa_{\bw}}+
	\left(\frac{d_{c}^\tL}{\nu_{c}^\tL}\right)^{\kappa_{\bw}}  \nonumber \\
	&= \left(\frac{\min\{|r_s^{\ast}-r_t^{\ast}|, n_r- |r_s^{\ast}-r_t^{\ast}|\}}{\nu_{r}^\tL}\right)^{\kappa_{\bw}}
	+\left(\frac{\min\{|c_s^{\ast}-c_t^{\ast}|, n_r- |c_s^{\ast}-c_t^{\ast}|\}}{\nu_{c}^\tL}\right)^{\kappa_{\bw}},
\end{align}
where $\bnu_{\bw}=(\nurL, \nucL)\tran$ and $0<\kappa_{\bw}\leq 1$.

We want to point out that since the labeling schemes of the physical and logical spatial structures are different, $Z_{\bv}$ and $Z_{\bw}$ are not identical when model~(\ref{eq:generalModel}) is applied to the GPU data. For the physical layout, cabinets from left to right in a row are indexed 1-25 and rows are indexed 1-8 from ``top to bottom". While for the logical layout, cabinets that are directly connected are considered to be adjacent and will be indexed numerically ascending. Referring once again to Figure~\ref{fig:logcial connections}, we see an illustration of the physical and logical labeling scheme between the 8 rows in each column. We can see that cabinets in physical location row 1 are connected to row 2 then row 4 and so on with a ``loopback" cable at the end to row 1. If we take a cabinet located at the physical location $(r,c) = (3,2)$ as an example, then it's logical location is $(r^{\ast}, c^{\ast}) = (8,2)$. The elements in $Z_{\bv}$ and $Z_{\bw}$ need to be arranged correspondingly to map each GPU unit to the correct physical and logical location. Due the complexity of logical connections, in the real data analysis in Section~\ref{sec:data_analysis}, the column and row index of the cabinet is sorted according to the logical connectivity of the cabinets from 1 to 25 and 1 to 8 but not the physical locations of the cabinets.

\subsection{The Positive Definiteness of Torus Correlation Matrix}\label{sec:dist.functions}

As noted by \shortciteN{gneiting2013strictly}, distances defined on a sphere can result in a non-positive definite correlation matrix. Our distance is different from that used in existing work because it is on a torus. Thus, some special attention is needed to ensure positive definiteness.

\begin{theorem}
When $0<\kappa_{\bw}\leq 1$, the correlation matrix defined on torus grid,
\begin{align}\label{eqref.bw.corr}
R_{\bw} = (\rho_{st}^{\tL}), \text{ where, } \rho_{st}^{\tL}= \exp\left[ - b^\tL_{st}(\nu_{\bw}, \kappa_{\bw}) \right], s=1,\ldots, m, t=1,\ldots, m.
\end{align}
is positive definite. Here, the distance $b^\tL_{st}(\nu_{\bw}, \kappa_{\bw})$ is defined in \eqref{eqn:logical.dist}.
\end{theorem}

Note that the correlation function is separable for its coordinates. That is
\begin{align*}
\rho_{st}^\tL = \exp\left[ - b^\tL_{st}(\nu_{\bw}, \kappa_{\bw}) \right]
=&\exp\left[-\left(\frac{\min\left\{|r_s^{\ast}-r_t^{\ast}|, n_r-|r_s^{\ast}-r_t^{\ast}|\right\}}{\nu_{r}^\tL}\right)^{\kappa_{\bw}}\right]
\\&\times \exp\left[-\left(\frac{\min\left\{|c_s^{\ast}-c_t^{\ast}|, n_c-|c_s^{\ast}-c_t^{\ast}|\right\}}{\nu_{c}^\tL}\right)^{\kappa_{\bw}}\right].
\end{align*}
In addition, a torus is the tensor product of two circles. Proposition~6.25 of \citeN{Wendland2004} proves a similar result that the tensor product of positive definite functions is also positive definite on Euclidean space using Fourier transformation. Because we consider the tensor product of two circles, we need to take a different approach to prove the theorem. We first sort the points on the torus grid as,
\begin{align}\label{eqn:torus.grid.sorted.location}
\left\{(r_1^{\ast}, c_1^{\ast}), (r_2^{\ast}, c_1^{\ast}), \dots, (r_{n_r}^{\ast}, c_1^{\ast}),\dots, (r_1^{\ast}, c_{n_c}^{\ast}), (r_2^{\ast}, c_{n_c}^{\ast}), \dots, (r_{n_r}^{\ast}, c_{n_c}^{\ast}) \right\},
\end{align}
which has $m$ locations in total from the $n_r\times n_c$ grid. Given the same column label $c^{\ast}$, the correlation matrix among those $n_r$ points is $A_{n_r\times n_r} = (\rho_{st}^{\tL})$, where
\begin{align}\label{eqn:w.sl.Amat}
\rho_{st}^{\tL}=\exp\left[-\left(\frac{\min\left\{|r_s^{\ast}-r_t^{\ast}|, n_r-|r_s^{\ast}-r_t^{\ast}|\right\}}{\nu_{r}^\tL}\right)^{\kappa_{\bw}}\right].
\end{align}
Note that the correlation function \eqref{eqn:w.sl.Amat} is equivalent to a correlation function defined on a circle. \citeN{gneiting2013strictly} showed that the powered exponential covariance matrix is positive definite for circles when $0<\kappa_{\bw}\leq 1$. Thus, $A$ is a positive definite matrix.

Similarly, given the same row label $r^{\ast}$, the correlation matrix among those $n_c$ points is $B_{n_c\times n_c} = (\rho_{st}^{\tL})$, where
\begin{align}\label{eqn:w.sl.Bmat}
\rho_{st}^{\tL}=\exp\left[-\left(\frac{\min\left\{|c_s^{\ast}-c_t^{\ast}|, n_c-|c_s^{\ast}-c_t^{\ast}|\right\}}{\nu_{c}^\tL}\right)^{\kappa_{\bw}}\right],
\end{align}
and $B$ is a also a positive definite matrix.

Let $\otimes$ be the Kronecker product, $\odot$ be the Hadamard (element-wise) product, and $J$ be a matrix of 1's. Based on the sorted locations in \eqref{eqn:torus.grid.sorted.location} and the separability of the correlation function, the correlation matrix for all $m$ locations is
\begin{align*}
R_{\bw} & = (J_{n_c\times n_c}\otimes A_{n_r\times n_r}) \odot (B_{n_c\times n_c}\otimes J_{n_r\times n_r})\\
& = (J_{n_c\times n_c}\odot B_{n_c\times n_c}) \otimes (A_{n_r\times n_r}\odot J_{n_r\times n_r})=B\otimes A.
\end{align*}
Because the eigenvalues of both $A$ and $B$ are all positive, the eigenvalues of $B\otimes A$ are the products of $A$'s and $B$'s, which are all positive. This result is shown in Theorem 8.5 of \citeN{Schott2016}. Thus, $B\otimes A$ is positive definite and so is $R_{\bw}$, proving the theorem.

\section{Bayesian Inference}\label{sec:Bayesian.inference}

The Bayesian framework is suggested for inference under the proposed model. Posterior sampling for the model is well within the capabilities of current Markov chain Monte Carlo (MCMC) algorithms providing a clear approach for inference. Other inferential methods, such as the EM algorithm, are available as alternatives, but will require additional time spent in derivation.

\subsection{Likelihood and Prior}

In the context of a Bayesian analysis, note the following relationship:
\[ \pi(\btheta|\bt,\boldsymbol{\delta}) \propto \mathcal{L}(\btheta|\bt,\boldsymbol{\delta})\pi(\btheta). \]
Here $\pi(\btheta)$ defines a prior distribution on the unknown parameters. The priors should provide an accurate representation of uncertainty in the parameters before data is observed. $\pi(\btheta|\boldsymbol{t},\boldsymbol{\delta})$ is the posterior distribution which updates the uncertainty in the parameters according to the observed data. For our proposed mixed modeling approach we define:
\begin{align*}
\btheta=
\left( \boldsymbol{\beta}\tran, \boldsymbol{v}\tran,  \boldsymbol{w}\tran, \boldsymbol{\sigma}\tran\right)\tran.
\end{align*}
where $\boldsymbol{\sigma}$ is a vector containing the parameters that determine the covariance structure.

For most analyses, the following independent specification is appropriate:
\[ \pi(\btheta) = \pi(\boldsymbol{\beta})\pi(\boldsymbol{v},\boldsymbol{w}|\boldsymbol{\sigma})\pi(\boldsymbol{\sigma}). \]
The vector, $\btheta$, contains the parameters of our model. This prior does not encode any dependence relationship between the mean and covariance parameters. However, this behavior is not preserved in the posterior. The vector of covariance parameters, $\boldsymbol{\sigma}$, is a set of hyperparameters for the random effects, therefore, random effects cannot be independent of $\boldsymbol{\sigma}$ \textit{a priori}. Note that $\boldsymbol{\sigma}$ also contains the error scale parameter, $\sigma$.

In the inference performed in this work, we avoid the use of improper priors, in order to ensure the existence of the posterior distribution. The amount of certainty in the prior needed to facilitate convergence of MCMC chains varies by parameter. A normal prior with large variance is typically an appropriate choice for $\pi(\boldsymbol{\beta})$. We suggest gamma priors for the positive covariance parameters. The length-scale parameters, $\nu^P_r,\nu^P_c,\nu^L_r,\nu^L_c$, typically require the most informative priors. Ultimately, the choice of prior distribution is left to the analyst. The priors we use for analyzing the GPU dataset can be found in Section \ref{sec:data_analysis}. The density, $\pi(\boldsymbol{v},\boldsymbol{w}|\boldsymbol{\sigma})$, is fully defined as a multivariate normal density in Section \ref{sec:data_model}.

We can now turn our attention to likelihood specification. The vector $\boldsymbol{\delta} = (\delta_1,  \delta_2,  \dots,  \delta_n )\tran$
contains the censoring information. Typically this is a one-hot encoding for observed failure, with a zero signifying right-censorship as defined in Section \ref{sec:data_model}. While we focus on right-censoring, this approach could be extended to handle left- and interval-censoring if the correct likelihood contributions are used. The vector $\bt$ is our observed event-time vector.

We now consider $\mathcal{L}(\btheta|\bt,\boldsymbol{\delta})$ which is the likelihood function. Under the independence assumptions,
\begin{align*}
 \mathcal{L}(\btheta|\bt,\boldsymbol{\delta}) = \prod_{i=1}^n \mathcal{L}_i(\btheta|t_i,\delta_i).
\end{align*}
This construction relies on the data being conditionally independent given the random effects. We can define the likelihood as,
\begin{align*}
\mathcal{L}_i(\btheta|t_i,0) = 1-\Phi\left(\frac{t_i-\mu_i}{\sigma}\right), \: \mathcal{L}_i(\btheta|t_i,1) = \frac{1}{\sigma t_i}\phi\bigg(\frac{t_i-\mu_i}{\sigma}\bigg),
\end{align*}
where $\Phi(\cdot)$ is the standard cumulative distribution function (cdf) and $\phi(\cdot)$ is the standard probability density function (pdf) of the location-scale distribution, $\FDIST$, defined in Section \ref{sec:data_model}. Note that standard implies location 0 and scale 1. To use the
lognormal distribution, one can replace $\Phi(\cdot)$ and $\phi(\cdot)$ with
$\Phi_{\text{NOR}}(\cdot)$ and $\phi_{\text{NOR}}(\cdot)$, which are the standard normal cdf and pdf, respectively. To use the Weibull distribution, one can use
\begin{align*}
\Phi_{\text{SEV}}(z)=1-\exp[-\exp(z)], \quad  \text{and} \quad
\phi_{\text{SEV}}(z)=\exp[z-\exp(z)],
\end{align*}
which are the cdf and pdf of the standard SEV distribution, respectively.
In this specification of the likelihood, we let $\boldsymbol{\mu} = (\mu_1, \mu_2, \dots, \mu_n)\tran$, where
\begin{align*}
\mu_i = \boldsymbol{x}_i\tran\boldsymbol{\beta} + \boldsymbol{z}_{vi}\tran\boldsymbol{v} + \boldsymbol{z}_{wi}\tran\boldsymbol{w}.
\end{align*}
The vector, $\boldsymbol{\mu}$ is the location parameter for $\FDIST$. Taken together with $\boldsymbol{\mu}$, which already has induced dependence through the random effects, the scale parameter, $\sigma$, defines the scale matrix of $\FDIST$.

\subsection{Posterior Sampling}

A sample from the posterior distribution is the most common approach for conducting posterior inference. These posterior draws are typically sampled through MCMC approaches. The software package, Stan, was used to produce the posterior chains presented in the remainder of this work. Stan requires full specification of the prior, likelihood and data. Stan utilizes the no-U-turn sampler (NUTS) and Hamiltonian Monte Carlo (HMC) algorithms. These algorithms provide a relatively efficient sampling approach when considering the time investment needed to self-code a sampler for the model proposed in this work.

When conducting posterior inference we rely on the posterior mean. Allow $\theta_{ps}, s=1,2,\dots,$ $M$, to be draws of parameter element $p$ in $\btheta$ from its posterior. Then the posterior mean can be estimated as,
\[ \frac{1}{M}\sum_{s=1}^M\theta_{ps}. \]

In the presence of outliers in the posterior chain, the posterior median is recommended as a point estimate for the unknown parameter. Draws from the posterior distribution, through MCMC chains, also allow for ease in access to uncertainty quantification. For an estimated $100(1-\alpha)\%$ Bayesian credible interval, the empirical $\alpha/2$ and $1-(\alpha/2)$ quantiles can be taken as interval bounds. Finally, we note that posterior draws allow for joint inference should that be of interest to the analyst.

\section{Simulation Study}\label{sec:simulation.study}

A simulation study was conducted in order to evaluate the performance of our proposed modeling approach.

\subsection{Simulation Design}

To simplify the simulation setting, the data was generated using a single four-level factor for the fixed effect structure. GPUs were evenly assigned to the four levels of our factor. The magnitudes of the parameters associated with this fixed effect structure were chosen to be comparable to estimates from the GPU dataset.

In all cases, a powered exponential covariance structure was used to model the correlation. For one set of random effects, distances were measured using our physical distance metric. For the other set of random effects, the logical distance metric was utilized. This mirrors the correlation structure used for modeling the GPU data.

In simulation, the number of locations varied. The three settings were 25, 49 and 100 locations; which allowed for an equal number of rows and columns in each case. In addition, we varied the number of replicates (i.e., the number of GPUs) at each location. We considered both 52 and 152 replicates. There were 50 unique datasets simulated for each simulation settings. All simulated datasets used the same hyperparameters, however the value of the random effects differed between simulations.

The empirical censoring rate was approximately 50\% but would vary slightly between simulated datasets. The censoring rate ensured that some failures were observed at each location for all settings and that parameters were estimable. Priors for the simulation are included in Table~\ref{tab:prior_sim}. These priors were chosen to facilitate convergence on the smaller scale seen in the simulation. Less informative priors were used in the full data analysis.

\begin{table}
\normalsize
  \begin{center}
    \caption{Prior specification for the model used in simulation. }
    \begin{tabular}{c|c|c|c}
    \hline
    \hline
      Parameter  &
      Prior &
      Parameter &
      Prior\\
      \hline
$\boldsymbol{\beta}$  & $\NOR(0,\sqrt{500}I)$ & $\lambda_v$  & $\GAMMA(1, 2) $ \\
$\sigma$  & $\GAMMA(0.1, 0.1) $  & $\sigma^2_w$ & $\GAMMA(4, 100)$    \\
$\sigma^2_v$  & $\GAMMA(6, 100)$  & $\nu_{r}^L$  & $\GAMMA(15, 10)$  \\
$\nu_{r}^P$ & $\GAMMA(9.5, 10)$ & $\nu_{c}^L$ & $\GAMMA(6, 10)$  \\
$\nu_{c}^P$ & $\GAMMA(9.2, 10)$ & $\kappa_w$  & $\BETA(18, 2)$  \\

     \hline
     \hline

   \end{tabular}
    \label{tab:prior_sim}
  \end{center}
\end{table}

Our primary metric of evaluating the performance of the model in simulation is the root mean-squared error (RMSE). We define RMSE as,
\[ \RMSE(\widehat{\theta}) = \sqrt{\sum_{q=1}^N\frac{(\widehat{\theta}_q-\theta)^2}{N}},\]
where $\widehat{\theta}$ is the chosen estimator for the unknown parameters and $\widehat{\theta}_q$ is the estimate calculated from the $q$th simulated dataset. In most cases, the posterior mean will be an appropriate choice of estimator. We use the posterior mean as our point estimator in this simulation study.

\subsection{Simulation Results}

\begin{figure}
	\centering
	\includegraphics[width=.85\textwidth]{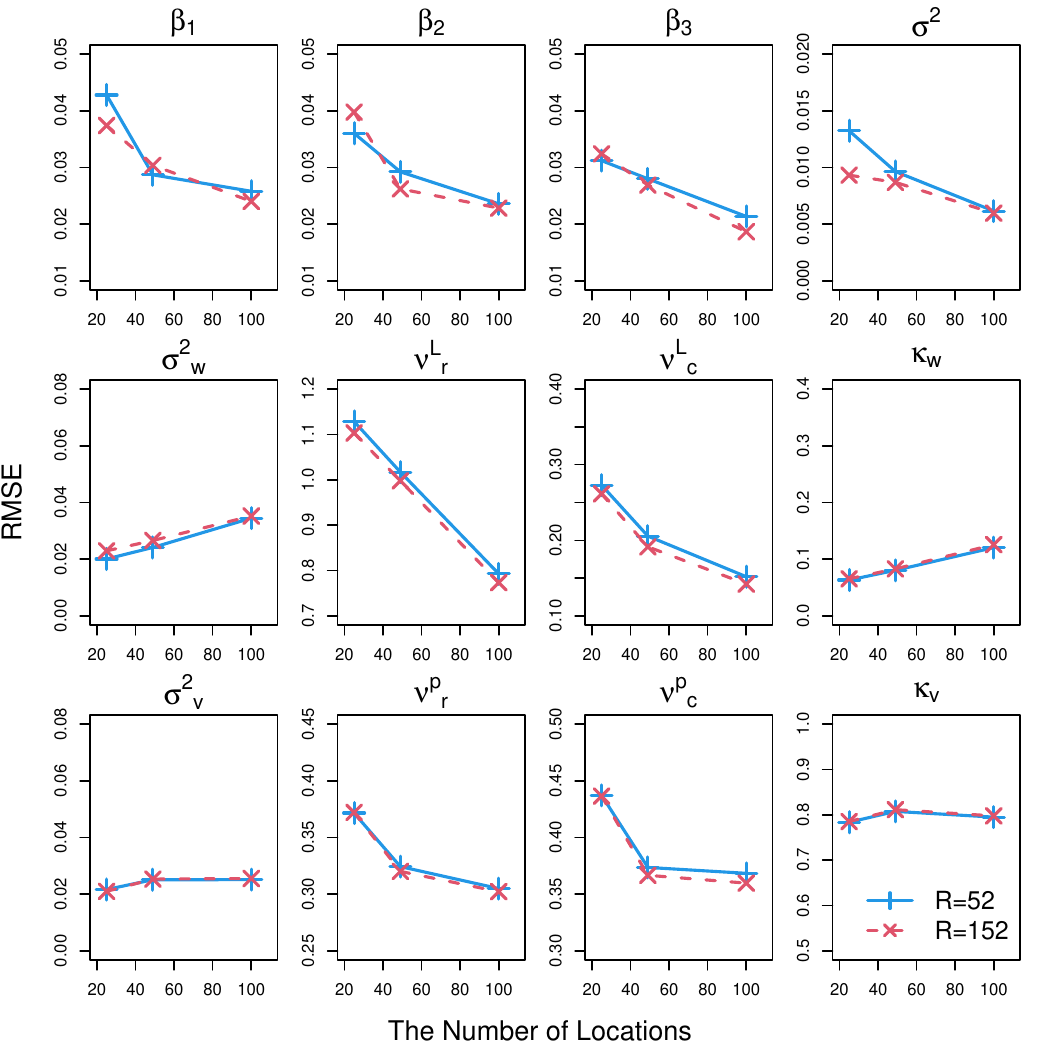}
	\caption{RMSE in estimation plotted against number of locations. Here $R$ is the number of GPUs in each location. }\label{fig:param_mse}
\end{figure}

Figure \ref{fig:param_mse} displays the RMSE plotted against the number of locations. Runs with 52 replicates are shown in solid blue lines while runs with 152 replicates are shown in dashed red lines. In general, the RMSE is decreasing with number of locations. In terms of fixed effect coefficients, this makes sense as the number of observations is increasing with the number of locations. This also makes sense with the correlation structure parameters since the number of random effects increases linearly with the number of locations which means there are more terms available to estimate variances and correlations.  Some trends that are not clearly decreasing are seen with the covariance parameters. However, these RMSE values are relatively small and there is also no obvious increase with the number of locations.

When looking at the fixed effects, we see a decrease in RMSE when increasing the number of replicates. With respect to the covariance function, the trend is less apparent. Note that increasing the number of locations increases the number of random effects, which allows for better estimation of the correlation function. On the other hand, increasing the number of replicates does not increase the number of random effects. Increasing the number of replicates can only decrease RMSE by decreasing the uncertainty in estimation of the estimated random effects.

Overall, the simulation suggests that the model is performing well. Estimation is quite accurate with 25 locations for many parameters. For some parameters, such as $\kappa_{\boldsymbol{w}}$, there is not clear improvement, in terms of estimation, as the number of locations increases. With more informed priors, we expect to see better performance in the posterior estimators.

\section{Data Analysis}\label{sec:data_analysis}
In this section, we apply the proposed method to the Cray XK7 Titan GPU dataset from~\shortciteN{ostrouchov2020gpu} as mentioned in Section~\ref{sec:data_model}. We start by reviewing the dataset in Section~\ref{sec:data_description}. We then perform exploratory data analysis in Section~\ref{sec:eda} and we close the section by presenting the results from the proposed mixed effects model in Section~\ref{sec:results}.

\subsection{GPU Data Description}\label{sec:data_description}
We start with a brief summary of the dataset we used in this section. As aforementioned, the GPU data contains the time-to-event information of the Titan supercomputer for over 6 years.  The physical distance and logical connections between GPUs play important roles in understanding GPU failure mechanisms. Specifically, for the Titan supercomputer, GPU nodes are placed in an $8 \times 25$ grid of cabinets. There are 3 cages within each cabinet, 8 slots within each cage, and 4 nodes within each slot. Each GPU was installed in a single node. Figure  \ref{fig:gpu}  shows the physical structure of the Titan supercomputer.  We focus on the off-the-bus (OTB) failure mode of the GPUs. In total, there are $n = 19,319$ units and $m = 200$ spatial locations.

There are a number of other important features of the GPU dataset. First, the data is split into ``old batch'' and ``new batch'' GPUs. It is believed that the failure processes of the old and new batches differ. Furthermore, there are many more failures recorded for the old batch GPUs. Therefore, our analysis will focus soley on the old batch data in order to avoid large biases in estimation. However, the results of the provided analysis are thought to be valid in the communication of relative effects of GPU placement within a cabinet (the fixed effects structure), and our spatial components.

Another important feature of the GPU data are the multiple failure modes. GPUs could experience off-the-bus (OTB) or double-bit error (DBE) failures. An OTB failure indicates a failure in communication between the CPU and GPU. A DBE results from a double-bit flip detection. Once a GPU has experienced one type of failure, the other failure type cannot be experienced. A competing risks model was previously investigated with a simpler spatial structure (\shortciteNP{min2023spatially}). Since the spatial structure is the main focus of this line of research, we will focus on the OTB failure distribution. GPUs that experience a DBE failure are right-censored.

When fitting the model, the row and column positions of each GPU are considered the location information to compute the physical and logical distance. The node, slot, and cage information are considered covariates that can affect a GPU's lifetime.

\subsection{Exploratory Data Analysis}\label{sec:eda}

\begin{figure}
\centering
\begin{minipage}{.5\textwidth}
  \centering
  \includegraphics[width=1\linewidth]{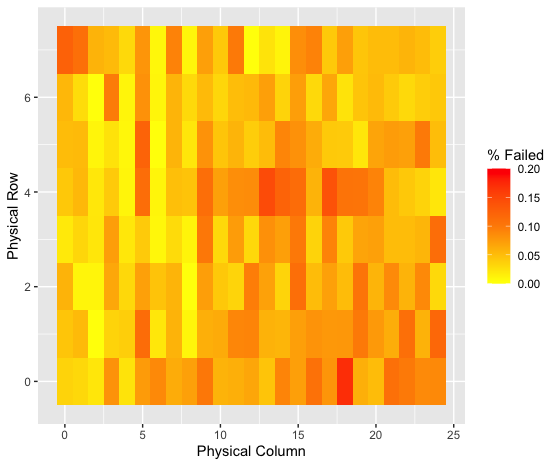}
\end{minipage}%
\begin{minipage}{.5\textwidth}
  \centering
  \includegraphics[width=1\linewidth]{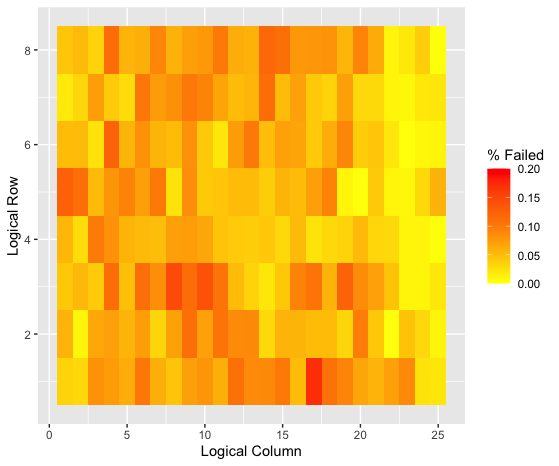}
\end{minipage}
\caption{ Failure rates based on physical location (left panel) and censoring rates based on logical location (right panel).}
\label{fig:censor}
\end{figure}

Before applying the proposed model to the GPU data, we conduct the exploratory analysis to better understand the data structure and assess how covariates impact the GPU lifetime. We would first like to understand the censoring structure of the dataset. Overall, we observe a censoring rate of 94.16\% across the entire dataset. The left panel of Figure \ref{fig:censor} displays the censoring rate at each location within the room. Notice that the censoring rate does vary substantially by location, with some locations seeing failure rates greater than 15\%. We note that the rate of observed failures is greatest for higher labeled columns and lower labeled rows.

It is also worth plotting the failure rate according to logical position. The right panel of Figure \ref{fig:censor} shows the same censoring rates but with locations reordered by their logical indices. In this case we still see high rates of correlation in the censoring rate, with the highest rates of censoring found at high column index.

\begin{figure}
	\centering
	\includegraphics[width=.7\textwidth]{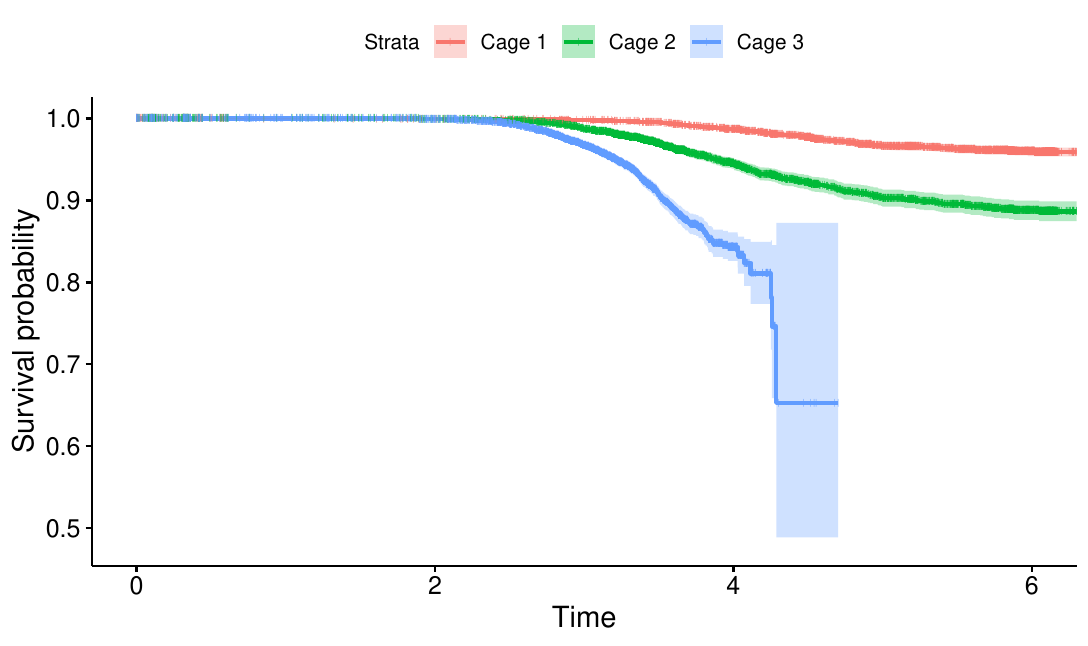}
	\caption{Kaplan-Meier estimates for the survival curve, stratified by cage. Cage 3, which is located highest in the cabinet experiences failures at the highest rate. 95\% confidence interval bounds are indicated by the shaded region.}\label{fig:km}
\end{figure}

For the purpose of modeling, we assume that the censoring mechanism is the same for the entire dataset. That is, censoring times are not a function of the location or fixed effect structure. We might also be interested in results from more traditional data analysis approaches. Figure \ref{fig:km} shows the Kaplan-Meier estimates separated by cage. Here, we note the high rate of censoring for each cage. We see a higher rate of failure for GPUs located in cage 3. These are the GPUs located at the highest point in the cabinet, farthest away from the cooling unit. This conclusion is supported by a p-value less than 0.01 for the log-rank test, suggesting a difference in survival curves.

\begin{figure}
\centering
\begin{tabular}{c}
\includegraphics[width=.7\linewidth]{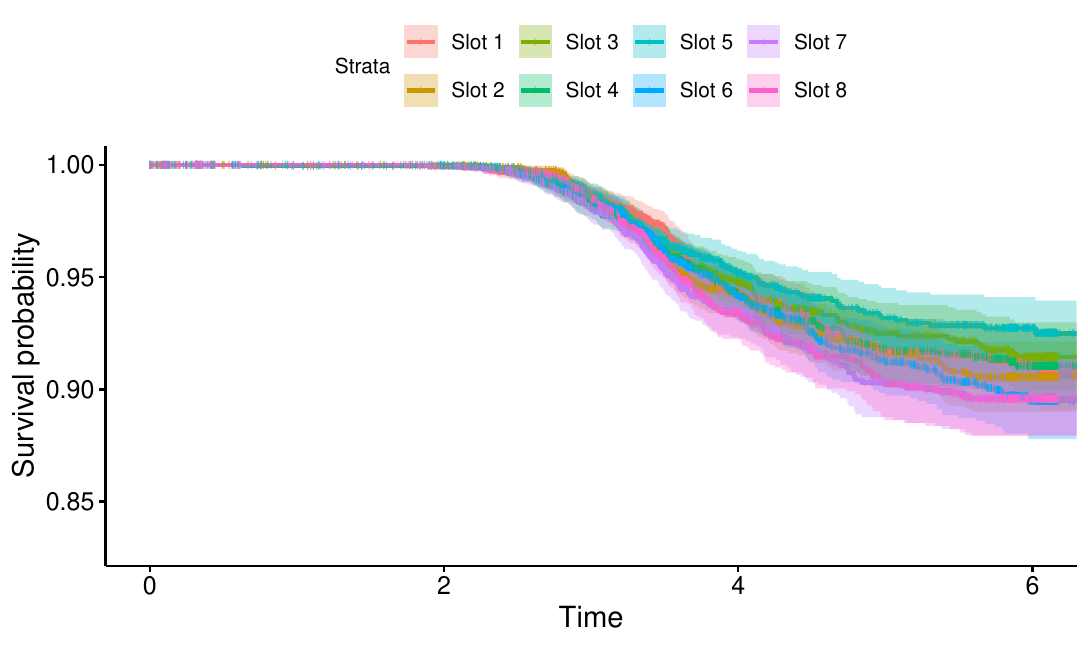}\\
\includegraphics[width=.7\linewidth]{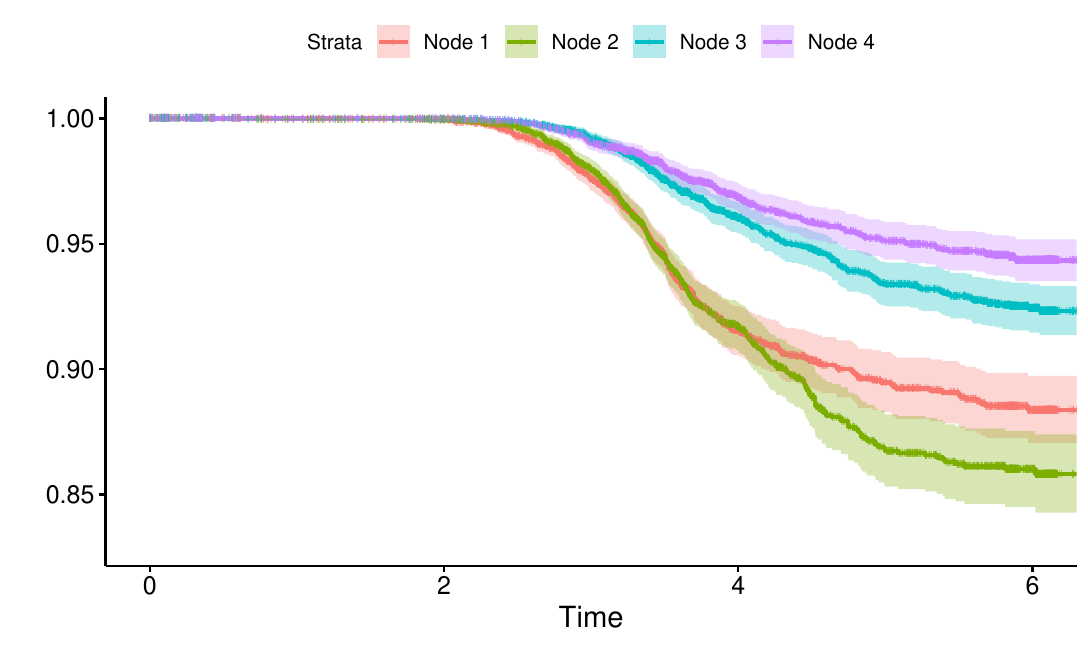}
\end{tabular}
\caption{Kaplan-Meier estimates stratified by slot (top panel) and Kaplan-Meier estimates stratified by node (bottom panel). }
\label{fig:slot_node}
\end{figure}

We also look at the Kaplan-Meier estimates when stratified by slot and node, shown in Figure \ref{fig:slot_node}. In the case of node, we see a p-value less than 0.01. However, when considering slot, the p-value is approximately 0.08. At the 0.05 significance level, there is not enough evidence to conclude that there are differences in the survival curves across level of slot.

\subsection{Analysis Results}\label{sec:results}

This section presents the analysis results from the proposed model. The priors used in our analysis are provided in Table~\ref{tab:prior}. For the fixed effect coefficients we used relatively diffuse normal priors. The correlation length parameters were given more informative priors in order to aid in convergence of the chain. In addition, we also apply a transformation to $\kappa_v$, using
\[ \kappa_v = \frac{2}{1+\exp(-\lambda_v)}. \]
The new parameter, $\lambda_v$, was introduced to simplify prior specification and to provide a parameter with a more desirable support. This will place 0 prior probability on $\kappa_v$ parameters less than 1. We also assume a normal distribution for the error vector, $\boldsymbol{\epsilon}$, when applying the mixed effects model to the GPU dataset. The included posterior inference is based off a single chain of 4{,}000 draws after a 4{,}000 draw burn-in. Posterior diagnostics suggest convergence.

\begin{table}
\normalsize
  \begin{center}
    \caption{Prior specification for the model used in data analysis. }
    \begin{tabular}{c|c|c|c}
    \hline
    \hline
      Parameter  &
      Prior &
      Parameter &
      Prior\\
      \hline
$\boldsymbol{\beta}$  & $\NOR(\zerovec,500I)$ & $\lambda_v$  & $\GAMMA(1, 2) $ \\
$\sigma$  & $\GAMMA(0.1, 0.1) $  & $\sigma^2_w$ & $\GAMMA(0.1, 0.1)$    \\
$\sigma^2_v$  & $\GAMMA(0.5, 0.5)$  & $\nu_{r}^L$  & $\GAMMA(2, 2)$  \\
$\nu_{r}^P$ & $\GAMMA(2, 2)$ & $\nu_{c}^L$ & $\GAMMA(5, 5)$  \\
$\nu_{c}^P$ & $\GAMMA(5, 5)$ & $\kappa_w$  & $\BETA(0.5, 0.5)$  \\

     \hline
     \hline

   \end{tabular}
    \label{tab:prior}
  \end{center}
\end{table}

\begin{table}
  \begin{center}
    \caption{Posterior results for the fixed effect structure. Cage 3 experiences the quickest rate of failure. Nodes 1 and 2 seem to be similar while nodes 3 and 4 are similar in terms of failure rate.}
    \begin{tabular}{ccrcrr}
    \hline
    \hline
    \multirow{ 2}{*}{Name} &
    \multirow{ 2}{*}{Parameter} &
    \multirow{ 2}{*}{Mean} &
    \multirow{ 2}{*}{SD} &
    \multicolumn{ 2}{c}{95\% Interval}\\\cline{5-6}
    &&&& Lower & Upper \\
      \hline
    Intercept & $\beta_0$ & 2.002 & 0.030 & 1.947 & 2.060 \\\hline
    Cage 1 & $\beta_1$ & 0.647 & 0.017 & 0.615 & 0.679 \\
    Cage 2 & $\beta_2$ & 0.263 & 0.015 & 0.234 & 0.293 \\\hline
    Slot 1 & $\beta_3$ & 0.046 & 0.027 & $-$0.008 & 0.097 \\
    Slot 2 & $\beta_4$ & 0.039 & 0.027 & $-$0.017 & 0.088 \\
    Slot 3 & $\beta_5$ & 0.062 & 0.026 & 0.011 & 0.112 \\
    Slot 4 & $\beta_6$ & 0.052 & 0.026 & 0.002 & 0.101 \\
    Slot 5 & $\beta_7$ & 0.075 & 0.026 & 0.025 & 0.124 \\
    Slot 6 & $\beta_8$ & 0.019 & 0.026 & $-$0.034 & 0.069 \\
    Slot 7 & $\beta_9$ & 0.003 & 0.025 & $-$0.045 & 0.052 \\\hline
    Node 1 & $\beta_{10}$ & $-$0.272 & 0.020 & $-$0.311 & $-$0.231 \\
    Node 2 & $\beta_{11}$ & $-$0.302 & 0.020 & $-$0.311 & $-$0.231 \\
    Node 3 & $\beta_{12}$ & $-$0.063 & 0.020 & $-$0.102 & $-$0.025 \\

    \hline
    \hline
   \end{tabular}
       \label{tab:fixed}
  \end{center}

\end{table}

We start by considering the fixed effects structure. These results are presented in Table~\ref{tab:fixed}. Posterior summaries for the intercept are included, however this parameter is not independently estimable. Cage 3 was used as a baseline category. The coefficients associated with the indicators for cages 1 and 2 are positive, indicating a larger mean failure-time for GPUs in those cages. This strengthens the preliminary conclusions from the exploratory data analysis.

Taken together, there doesn't seem to be a slot effect, though the three middle-slot intervals do not include 0, which is consistent with possibly faster airflow near the middle of a cabinet. The node does seem to be related to failure-time, though the relationship is weaker than that between cage and failure-time.

\begin{table}
  \begin{center}
      \caption{Posterior results for the random effect structure. Since $\sigma^2_v > \sigma^2_w$ there appears to be more variability explained by the physical locations as opposed to the logical connections. Differing correlation lengths suggest an anisotropic structure.}
    \begin{tabular}{cccrr}
    \hline
    \hline
    \multirow{ 2}{*}{Parameter} &
    \multirow{ 2}{*}{Mean} &
    \multirow{ 2}{*}{SD} &
    \multicolumn{ 2}{c}{95\% Interval}\\\cline{4-5}
    &&& Lower & Upper \\
      \hline
    $\sigma^2$ & 0.182 & 0.009 & 0.166 & 0.200 \\
    $\sigma^2_v$ & 0.022 & 0.018 & 0.004 & 0.064 \\
    $\nu_{r}^P$ & 0.964 & 0.351 & 0.424 & 1.805 \\
    $\nu_{c}^P$ & 0.964 & 0.369 & 0.426 & 1.839 \\
    $\kappa_v$ & 1.297 & 0.174 & 1.016 & 1.635 \\
    $\sigma^2_w$ & 0.012 & 0.006 & 0.004 & 0.028 \\
    $\nu_{r}^L$ & 1.876 & 0.799 & 0.716 & 3.834 \\
    $\nu_{c}^L$ & 0.598 & 0.307 & 0.214 & 1.346 \\
    $\kappa_w$ & 0.945 & 0.073 & 0.743 & 1.000 \\

    \hline
    \hline

   \end{tabular}
    \label{tab:table}
  \end{center}
\end{table}

\begin{figure}
\centering
\begin{minipage}{.5\textwidth}
  \includegraphics[width=1\linewidth]{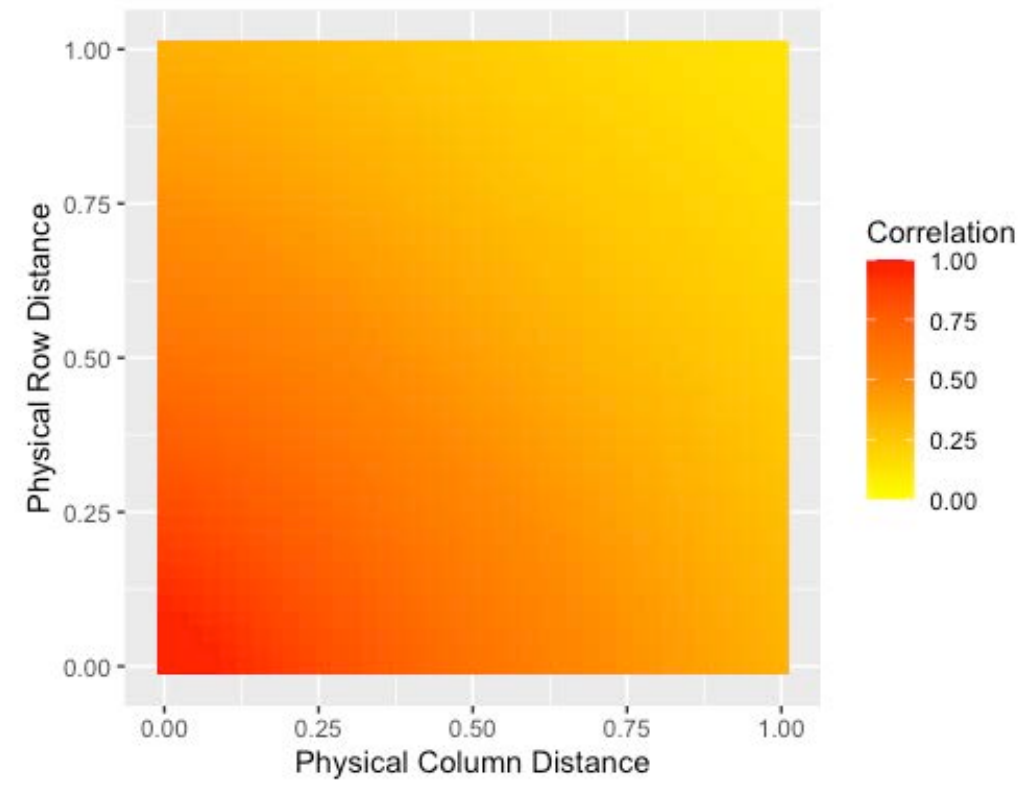}
\end{minipage}%
\begin{minipage}{.5\textwidth}
  \centering
  \includegraphics[width=1\linewidth]{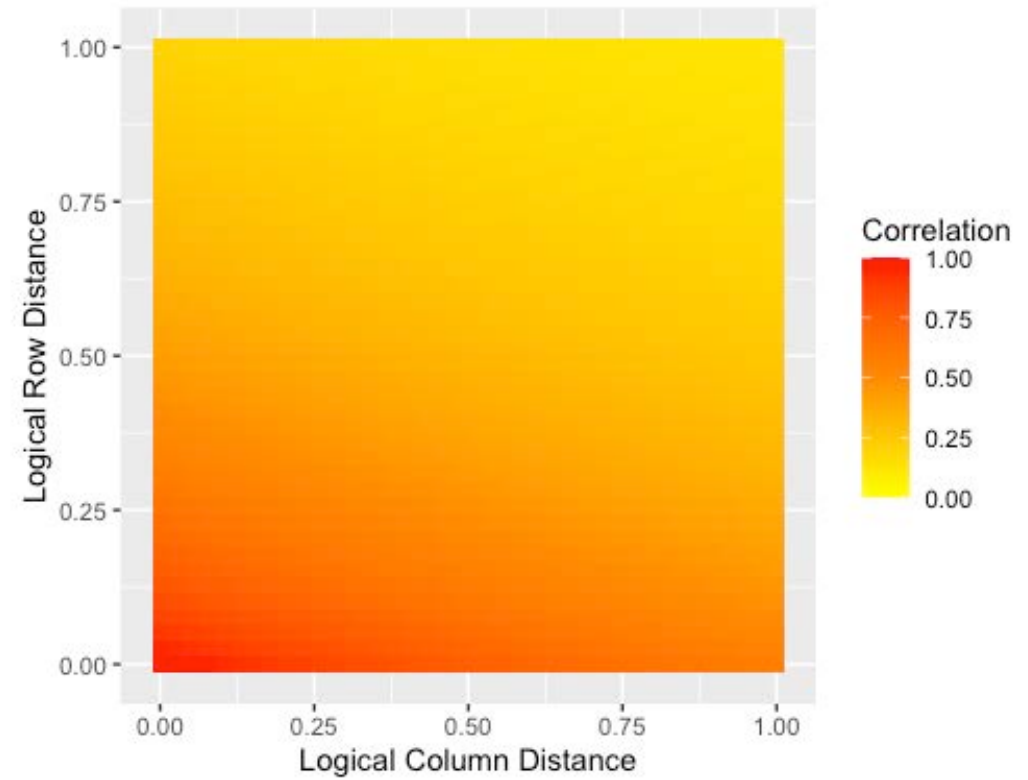}
\end{minipage}
\caption{Estimated correlation function based on physical distance (left panel) and estimated correlation function based on logical distance (right panel).}
\label{fig:estimated.corr.heatmap}
\end{figure}

We now focus on the covariance structure parameters. Figure~\ref{fig:estimated.corr.heatmap} illustrates the estimated correlation functions based on physical and logical distances. The physical distance exhibits a strong correlation, suggesting that room temperature may have a significant impact on failures. In making formal inference on the spatial components, we notice first that $\sigma^2_{\boldsymbol{v}} > \sigma^2_{\boldsymbol{w}}$ with posterior probability 0.725. This indicates that the physical location is a larger indicator of failure-time than the logical position with the random effects tending to be farther from zero.

When focusing on the correlations in physical space, the correlation length parameters appear to be similar for the two directions. However, when looking at the logical correlation structure, the correlation lengths differ dramatically, indicating a different decay speed in correlation when moving across columns and rows in logical space, as shown in the estimation of the logical correlation function in Figure~\ref{fig:estimated.corr.heatmap}. This indicates potential imbalances in job scheduling between columns and rows. Also of particular interest is the high amount of posterior probability close to 1 for $\kappa_{\boldsymbol{w}}$. This indicates that an exponential correlation structure is likely suitable for explaining the data.

With the complexity of the spatial structure, it is important to consider the possibility that a simpler model is appropriate. We use the Bayes factor as a means for comparing two models, $M_1$ and $M_0$. Bayes factor is defined in the following manner:
\[ \text{BF}_{10} = \frac{\int L(\boldsymbol{\theta}|\boldsymbol{y},M_1)\pi(\boldsymbol{\theta}|M_1)d\boldsymbol{\theta}}{\int L(\boldsymbol{\theta}|\boldsymbol{y},M_0)\pi(\boldsymbol{\theta}|M_0)d\boldsymbol{\theta}}. \]
Notice that $\text{BF}_{10} > 1$ is evidence in support of $M_1$, while $\text{BF}_{10} < 1$ is evidence in support of $M_0$.

In our comparison, we consider the following three models:
\begin{align*}
M_0 : & y_{ij} = \boldsymbol{x_i}\tran\boldsymbol{\beta} + \epsilon_{ij}, \\
M_1 : & y_{ij} = \boldsymbol{x}_i\tran\boldsymbol{\beta} + \boldsymbol{z}_i\tran\boldsymbol{v} + \epsilon_{ij}, \\
M_2 : & y_{ij} = \boldsymbol{x}_i\tran\boldsymbol{\beta} + \boldsymbol{z}_{vi}\tran\boldsymbol{v} + \boldsymbol{z}_{wi}\tran\boldsymbol{w} + \epsilon_{ij},
\end{align*}
with each model incorporating the appropriate priors. These are a model without random effects, $M_0$; a traditional spatial model, $M_1$; and the proposed spatial model with random effects defined by both physical and logical locations, $M_2$.

In comparing these models we obtain $\text{BF}_{01} \approx 0$ which indicates that there is essentially no evidence to prefer the fully fixed effects model to the traditional spatial model. Furthermore, $\text{BF}_{12} = 5.88\times10^{-6}$, indicates strong support for our proposed model. Taken together, this suggests that there is need for the inclusion of random effects defined with logical distances. The cable connections induce correlations in the Titan GPU dataset and this correlation structure needs to be considered in effective modeling efforts.

\section{Conclusions}\label{sec:conclusion}

In this paper, we develop a spatially correlated failure-time model under two types of spatial random components, motivated by the Titan GPU lifetime data.  In general, some time-to-failure datasets may contain complicated correlation structures. Identifying components of the correlation structure allows for separate modeling of the sources. The Kronecker product of positive definite matrices will be positive definite. This suggests that when valid correlation functions, like those defined on different spatial dimensions, are multiplied we define a correlation function that defines a positive definite covariance matrix.

The Titan GPU dataset served as an interesting motivating application for our modeling efforts. This dataset contained failure information on GPUs spread across 200 locations. The physical location of the GPUs were thought to be an important feature of the failure structure. Additionally, the number of wired connections between locations was also thought to be important, thus defining a logical distance metric. This interesting spatial structure motivated the modeling of two separate sets of random effects.

Our modeling efforts were validated through simulation. We focused on simulating data with a correlation structure similar to what is seen with the GPU data. We varied the number of locations across datasets. In doing so, we found that our proposed model produced accurate estimates of fixed effect parameters for datasets with as little as 25 locations. Furthermore, we see improved accuracy in estimation for many covariance structure parameters with the increasing number of locations. The simulation study suggested that use of our proposed model was appropriate.

There are a number of directions for extension of this work. While the logical connections seen in the GPU dataset provided a single interesting spatial structure for study, there are many more spatial structures that could be investigated in future modeling work. There are also further avenues for analyzing the GPU data. We note that a exponential covariance structure appears to be appropriate for the logical spatial random effects. Reducing the complexity of the model would facilitate convergence of MCMC chains, offering a promising direction for future work. The GPU dataset also contains information on a second mode of failure. The spatial modeling of this work could likely be extended to handle competing risks data.

\section*{Acknowledgments}
The authors acknowledge the Advanced Research Computing program at Virginia Tech for providing computational resources.
	




\end{document}